\documentclass[]{aa}
\usepackage[varg]{txfonts}
\usepackage{graphicx}
\usepackage{natbib,twoopt}
\usepackage[breaklinks=true]{hyperref} 
\bibpunct{(}{)}{;}{a}{}{,} 
\makeatletter
 \newcommandtwoopt{\citeads}[3][][]{\href{https://ui.adsabs.harvard.edu/abs/#3/abstract}%
 {\def\hyper@linkstart##1##2{}%
 \let\hyper@linkend\@empty\citealp[#1][#2]{#3}}}
 \newcommandtwoopt{\citepads}[3][][]{\href{https://ui.adsabs.harvard.edu/abs/#3/abstract}%
 {\def\hyper@linkstart##1##2{}%
 \let\hyper@linkend\@empty\citep[#1][#2]{#3}}}
 \newcommandtwoopt{\citetads}[3][][]{\href{https://ui.adsabs.harvard.edu/abs/#3/abstract}%
 {\def\hyper@linkstart##1##2{}%
 \let\hyper@linkend\@empty\citet[#1][#2]{#3}}}
 \newcommandtwoopt{\citeyearads}[3][][]%
 {\href{https://ui.adsabs.harvard.edu/abs/#3/abstract}
 {\def\hyper@linkstart##1##2{}%
 \let\hyper@linkend\@empty\citeyear[#1][#2]{#3}}}
\makeatother

\begin{document}

   \title{Activity of the Jupiter co-orbital comet P/2019~LD$_{2}$ (ATLAS)  observed with OSIRIS at the 10.4~m GTC\thanks{Based
          on observations made with the GTC telescope, in the Spanish 
          Observatorio del Roque de los Muchachos of the Instituto de 
          Astrof\'{\i}sica de Canarias (program ID GTCMULTIPLE2F-20A).}}
   \author{J. Licandro\inst{1,2}
           \and
           J. de Le\'on\inst{1,2}
           \and
           F. Moreno\inst{3}
           \and
           C. de la Fuente Marcos\inst{4}
            \and
           R. de la Fuente Marcos\inst{5}
           \and
           A. Cabrera-Lavers\inst{6,1,2}
           \and
           L. Lara\inst{3}
           \and
           A. de Souza-Feliciano\inst{7}
           \and
           M. De Pr\'a\inst{8}
           \and
           N. Pinilla-Alonso\inst{8}
           \and
           S. Geier\inst{6,1}
           }
   \authorrunning{Licandro et al.}
   \titlerunning{GTC observations of Jupiter co-orbital comet P/2019~LD$_{2}$ (ATLAS)}
   \offprints{J. Licandro, \email{jlicandr@iac.es}}
   \institute{Instituto de Astrof\'{\i}sica de Canarias (IAC), 
              C/ V\'{\i}a L\'actea s/n, E-38205 La Laguna, Tenerife, Spain
              \and
              Departamento de Astrof\'{\i}sica, Universidad de La Laguna, 
              E-38206 La Laguna, Tenerife, Spain
              \and
              Instituto de Astrof\'\i sica de Andaluc\'\i a, CSIC, Glorieta de la Astronom\'\i a s/n, 18008 Granada, Spain
              \and
              Universidad Complutense de Madrid,
              Ciudad Universitaria, E-28040 Madrid, Spain 
              \and
              AEGORA Research Group,
              Facultad de Ciencias Matem\'aticas,
              Universidad Complutense de Madrid,
              Ciudad Universitaria, E-28040 Madrid, Spain
              \and
              GRANTECAN, 
              Cuesta de San Jos\'e s/n, E-38712 Bre\~na Baja, La Palma, Spain
              \and
              Observat\'orio Nacional, Rio de Janeiro, 20921-400, Brazil
              \and
              Florida Space Institute, 12354 Research Parkway Partnership 1 Building, Suite 214 Orlando, FL 32826-0650, USA}
   \date{Received XXXX / Accepted XX Xxxxxxxx XXXX}

   \abstract
    {The existence of comets  with heliocentric orbital periods close to that of Jupiter  (i.e., co-orbitals) has been known for some time. Comet 295P/LINEAR (2002~AR$_2$) is  a well-known quasi-satellite of Jupiter. However, their orbits are not long-term stable, and they may eventually experience flybys with Jupiter at very close range, close enough to trigger tidal disruptions like the one suffered by comet Shoemaker-Levy~9 in 1992.     }
    {Our aim was to study the observed activity and the dynamical evolution of the Jupiter transient co-orbital comet P/2019~LD$_2$ (ATLAS) and its dynamical evolution.
     }
    {We present results of an observational study of P/2019~LD$_2$ carried out with the 10.4~m Gran~Telescopio~Canarias (GTC) that includes image analyses using a Monte Carlo dust tail fitting code to characterize its level of cometary activity, and spectroscopic studies to search for gas emission. We also present $N$-body simulations to explore its past, present, and future orbital evolution.
     }
    {  Images of P/2019~LD$_{2}$ obtained on 2020 May 16, show a conspicuous coma and tail, but the spectrum  obtained on 2020 May 17, does not exhibit any evidence of CN, C$_2$, or C$_3$ emission. The comet brightness in a 2.6\arcsec aperture diameter is $r'=19.34\pm0.02$~mag, with colors $(g'-r')=0.78\pm0.03$, $(r'-i')=0.31\pm0.03$, and $(i'-z')=0.26\pm0.03$.  The temporal dependence of the dust loss rate of P/2019~LD$_{2}$ can be parameterized by a Gaussian function having a full width at half maximum of 350 days, with a maximum dust mass loss rate of 60 kg s$^{-1}$ reached on 2019 August 15. The total dust loss rate from the beginning of activity until the GTC observation date (2020 May 16) is estimated at
1.9$\times$10$^{9}$ kg. Comet P/2019~LD$_{2}$ is now an ephemeral co-orbital of Jupiter, following what looks like a short arc of a quasi-satellite cycle that started in 2017 and will end in 2028. On 2063 January 23, it will experience a very close encounter with Jupiter at perhaps 0.016~au;  its probability of escaping the solar system during the next 0.5~Myr is estimated to be 0.53$\pm$0.03.
     }
    {Photometry and tail model results show that P/2019~LD$_{2}$ is a kilometer-sized object, in the size range of the Jupiter-family comets, with a typical comet-like activity most likely linked to sublimation of crystalline water ice and clathrates. Its origin is still an open question. Our numerical studies give a probability of this comet having been captured from interstellar space during the last 0.5~Myr of 0.49$\pm$0.02 (average and standard deviation), 0.67$\pm$0.06 during the last 1~Myr, 0.83$\pm$0.06 over 3~Myr, and 0.91$\pm$0.09 during the last 5~Myr.
     }


   \maketitle

   \section{Introduction\label{intro}}

Comet P/2019~LD$_{2}$ (ATLAS), hereafter LD$_{2}$, was discovered in early June 2019 by the Asteroid Terrestrial-impact Last Alert System (ATLAS; \citealt{2018PASP..130f4505T}) as a faint asteroidal object. It was initially classified as a Jupiter trojan. In-depth inspection of images obtained in 2019 revealed a faint tail, suggesting that it presented comet-like activity   (see \url{http://www.ifa.hawaii.edu/info/press-releases/2019LD2/}). This cometary nature was explicitly acknowledged with the publication of MPEC~2020-K134\footnote{https://www.minorplanetcenter.net/mpec/K20/K20KD4.html} and CBET~4780. Prediscovery observations made by the Dark Energy Camera (DECam) on 2018 August 10, were reported on 2020 August 2, with the publication of CBET~4821 \citep{2020CBET.4821....1S}. They also reported that the object was not detected in DECam images acquired in 2017, suggesting that activity started sometime between 2017 and 2018 and that  a compact coma was present at the time when the August 2018 images were taken. These observations are consistent with an upper limit of the nucleus radius of around 3~km.

As no signs of comet-like activity have been detected on a  Jupiter trojan, even though it is widely accepted that they are captured objects from the outer solar system  \citep[e.g.,][]{2005Natur.435..462M,2018NatAs...2..878N}, the discovery of persistent activity on a putative Jupiter trojan asteroid is a very important result as it may suggest that some of them contain volatile material (likely water ice) on their surfaces. They are originally expected to consist of a combination of rock, dust, ice, and frozen gases. However, so far no activity has been detected by sublimation of water ice or other volatiles in any of them, which would confirm this hypothesis. 

For this reason we scheduled observations of LD$_{2}$  with the world's largest optical telescope, the 10.4m Gran Telescopio Canarias (GTC), to study the possible comet-like activity of LD$_{2}$  as soon as it became visible in May 2020. We also started a numerical exploration of its dynamical properties to determine if it could be a true member of the primordial Trojan population.

In the meantime,  \citet{2020RNAAS...4...74K} used the orbital elements determined adding new astrometric data obtained in 2020 to show that the comet had experienced a close encounter with Jupiter on 2017 February 17, at 0.092~au, well inside the Hill radius of the planet (0.338~au). These authors concluded that LD$_{2}$ is a recently captured centaur, not a Jupiter trojan. As  the value of its Tisserand parameter relative to Jupiter is $T_{\rm J} = 2.94$, LD$_{2}$ can be classified as a Jupiter-family comet (JFC) according to \citet{1997Icar..127...13L}. Even if LD$_{2}$ is not a Jupiter trojan, it is certainly a very interesting object that could help us understand better the transition from centaur to JFC, the activation mechanisms of these bodies at large heliocentric distances, and how activity affects the surface of centaurs. LD$_{2}$ orbits the Sun just beyond Jupiter,  and its activity is similar to  that of 29P/Schwassmann-Wachmann-1, an object considered  a prototypical ``gateway''  between the centaurs and JFCs by \citet{2019ApJ...883L..25S}.

In this paper we present the results of the observations (visible images and spectra) obtained with the 10.4m GTC, and the dynamical properties of  LD$_{2}$ derived from the analysis of an extensive sample of $N$-body simulations. In Sect. \ref{obs} we describe the observations and data reduction, derive the absolute magnitude and colors obtained from the images, and present the gas production rate upper limits derived from the spectra. In Sect. \ref{model} we present the analysis of the activity  based on the observed dust tail using a Monte Carlo dust scattering model. In Sect. \ref{evolution}  we present the results of the $N$-body simulations, and describe the past, present, and future dynamical evolution of LD$_{2}$. Our conclusions are laid out in Sect. \ref{conclusions}.

   \section{Observations\label{obs}}
\begin{figure}
    \centering
     \includegraphics[width=\linewidth]{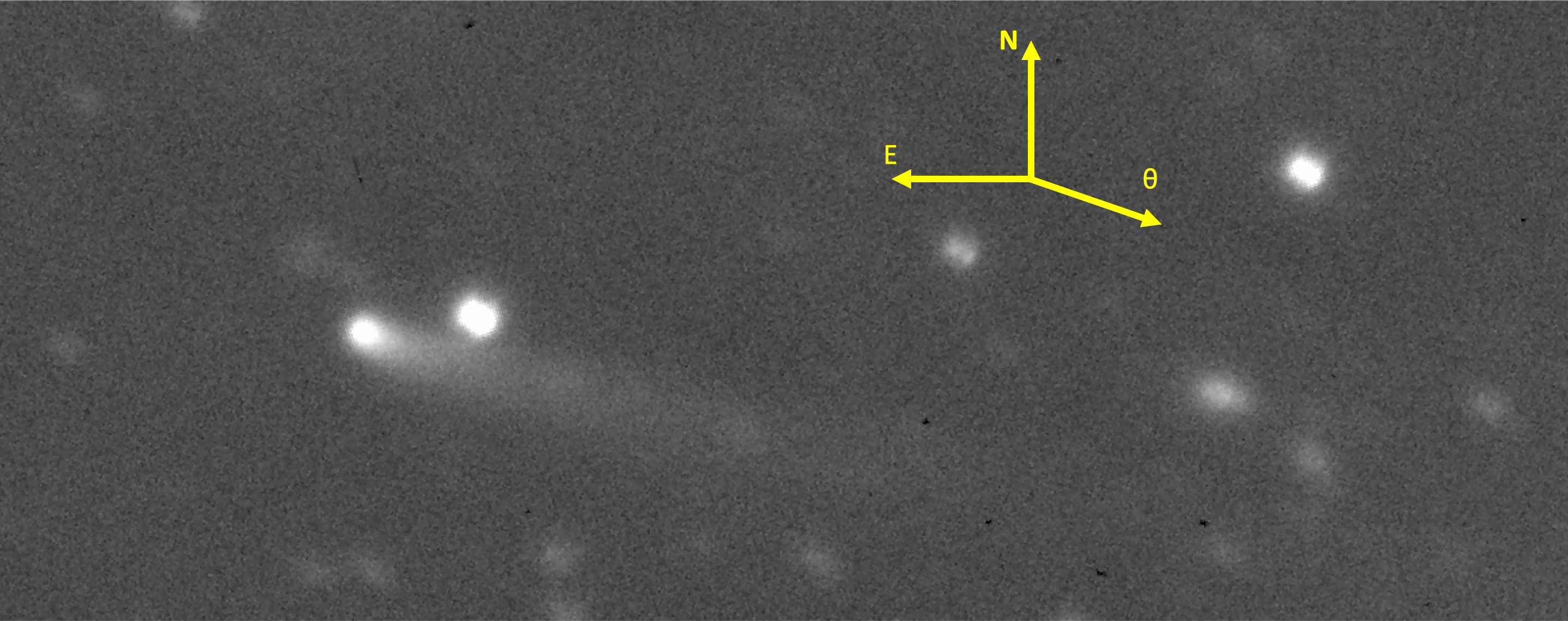}
     \caption{Image of P/2019~LD$_2$ (ATLAS) obtained on 2020 May 16. The image shown is the sum of four images, 60~s exposure time each, obtained using the $r'$-band filter. The field is 150\arcsec$\times$60\arcsec; north is up, east to the left. The object presents a conspicuous comet-like coma and tail almost aligned with the extended Sun-to-target radius vector ($\theta$).}
     \label{image}
\end{figure}

\begin{figure}
    \centering
     \includegraphics[width=\linewidth]{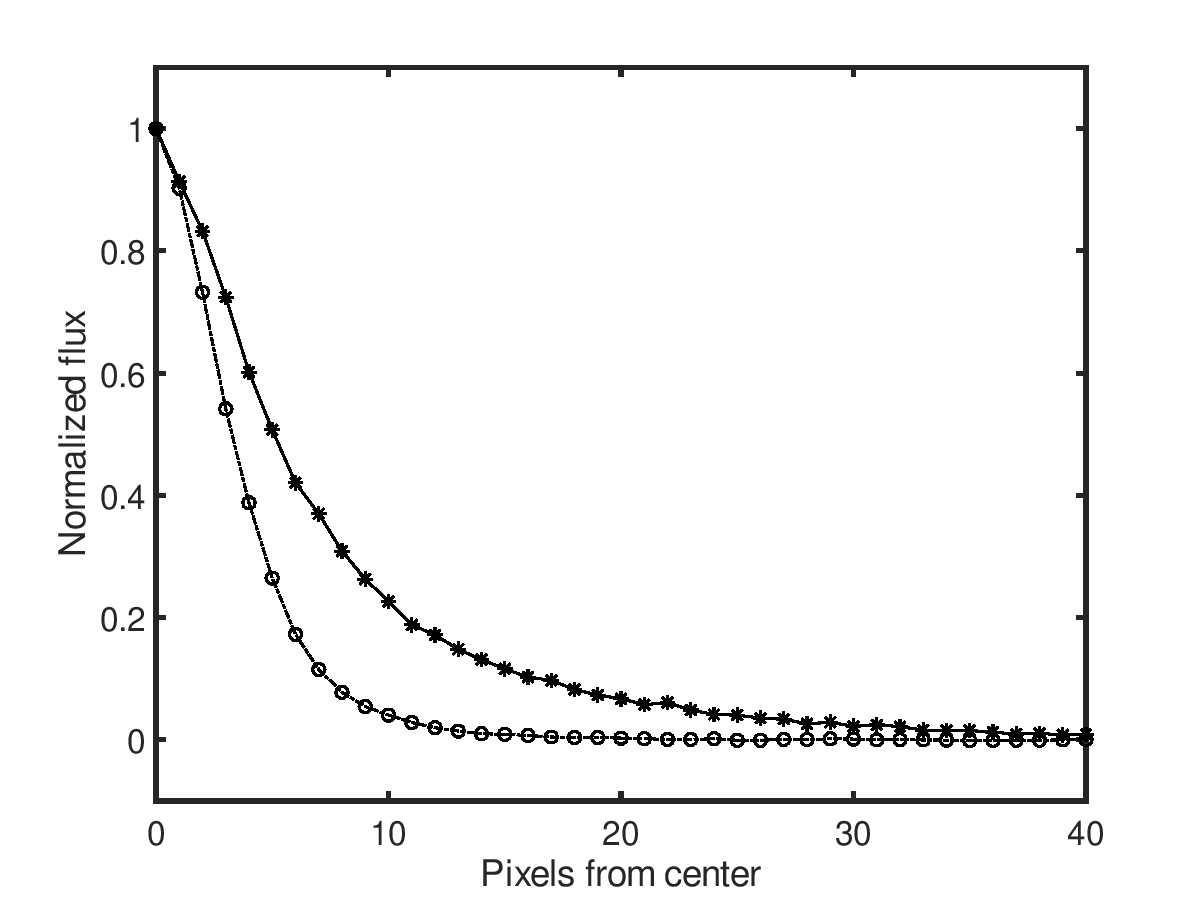}
     \caption{Brightness normalized radial profile of P/2019~LD$_2$ (ATLAS), plotted as stars, compared to the profile of a field star, plotted as open circles. The profiles were obtained from one single 60~s exposure time r-band image taken on 2020 May 16. The comet profile is wider than that of the field stars because of the presence of a conspicuous coma.}
     \label{perfil}
\end{figure}

We obtained CCD images of  LD$_{2}$ on  2020 May 16,  and low-resolution visible spectra on 2020 May 17, using the Optical System for Imaging and Low Resolution Integrated Spectroscopy (OSIRIS) camera-spectrograph \citep{2010ASSP...14...15C} at the 10.4m Gran Telescopio Canarias (GTC). The observational circumstances are shown in Table \ref{observations}. The OSIRIS detector is a mosaic of two Marconi 2048$\times$4096 pixel CCDs. The total unvignetted field of view is 7.8\arcmin$\times$7.8\arcmin, and the plate scale is 0.127~\arcsec/pix. Standard operation mode consists of a 2$\times$2 binning, with a readout speed of 200~kHz (with a gain of 0.95~e$^{\rm -}$/ADU and a readout noise of 4.5~e$^{\rm -}$). On May 16 we obtained individual images using the Sloan {\em g',r',i',z'} filters with individual exposure times of 60 seconds.  We did one {\em r', g', r', i', r', z', r'}  sequence of images with the  telescope tracking at the comet's proper motion. Images  were bias and flat-field corrected (using sky flats). The comet presents a conspicuous coma and tail as seen in Fig.~\ref{image}. The full width at half maximum (FWHM) of the point spread function (PSF) of  the comet, measured in one of the {\em r'} 60-second images, is wider than that of the stars, 2.6\arcsec versus 1.5\arcsec  ~(see Figure~\ref{perfil}) and the tail is $>1${\arcmin} long ($>$ 1.9 $\times$ 10$^{5} $~km at the comet distance). 

Aperture photometry was computed using  standard tasks in the Image Reduction and Analysis Facility (IRAF\footnote{IRAF is distributed by the National Optical Astronomy Observatory, which is operated by the Association of Universities for Research in Astronomy, Inc., under cooperative agreement with the National Science Foundation.}), following a procedure similar to that described in \citet{2019A&A...625A.133L}.  Using an aperture diameter equivalent to the comet's FWHM (2.6\arcsec), we obtained a  magnitude $r' = 19.34\pm0.02$ and the colors $(g'-r') = 0.78\pm0.03$, $(r'-i') = 0.31\pm0.03$, and $(i'-z')= 0.26\pm0.03$.  The background sky was measured (and subsequently subtracted) as the median value in a region close to the comet free of coma, tail, and background stars. Flux calibration was done using GTC zero-points computed for each observing night and provided by the telescope support astronomer. 
The colors of the comet, which correspond to the colors of the dust coma in the  aperture used, are much larger than the solar values, $(g'-r') = 0.44\pm0.02$, $(r'-i') = 0.11\pm0.02$, and $(i'-z')= 0.02\pm0.03$ (see https://www.sdss.org/dr12/algorithms/ugrizvegasun/), showing that the dust coma is redder than the Sun. 
 
 From the apparent magnitude we derived the absolute magnitudes in the $g$' filter using Eq. (1) from \citet{2019ApJ...886L..29J},  obtaining $H_g = 13.10\pm0.03$~mag. Assuming a visible geometric albedo between 0.1 and 0.04, this provides an upper limit for the nucleus radius $R_N$ between 5.0 and 8.0~km. Considering the conspicuous activity observed, the diameter of the comet nucleus should be much smaller than these values, in agreement with the \citet{2020CBET.4821....1S} results.



%
%
%
%

%
%

We also obtained two visible spectra of P/2019~LD$_{2}$ on  2020 May 17,  with the aim of looking for signatures of the typical gas species observed in comets. Each individual spectrum consisted of an exposure of 600 seconds using the R300B grism and the 1.49{\arcsec}~slit width, covering a wavelength range from 3600 to 7500~\AA, and with a dispersion of 4.96~\AA/pix for a 0.6{\arcsec}~slit.  As the aim was to look for gas species, the slit was oriented in the direction of the comet tail, not in parallactic angle. This allows a better study of the gas emission along the tail, but introduces a significant error in the slope of the spectrum due to the differential atmospheric refraction.  For this reason we did not use this spectrum to compute the spectral slope (color) of the comet. 

\begin{figure}
    \centering
     \includegraphics[width=\linewidth]{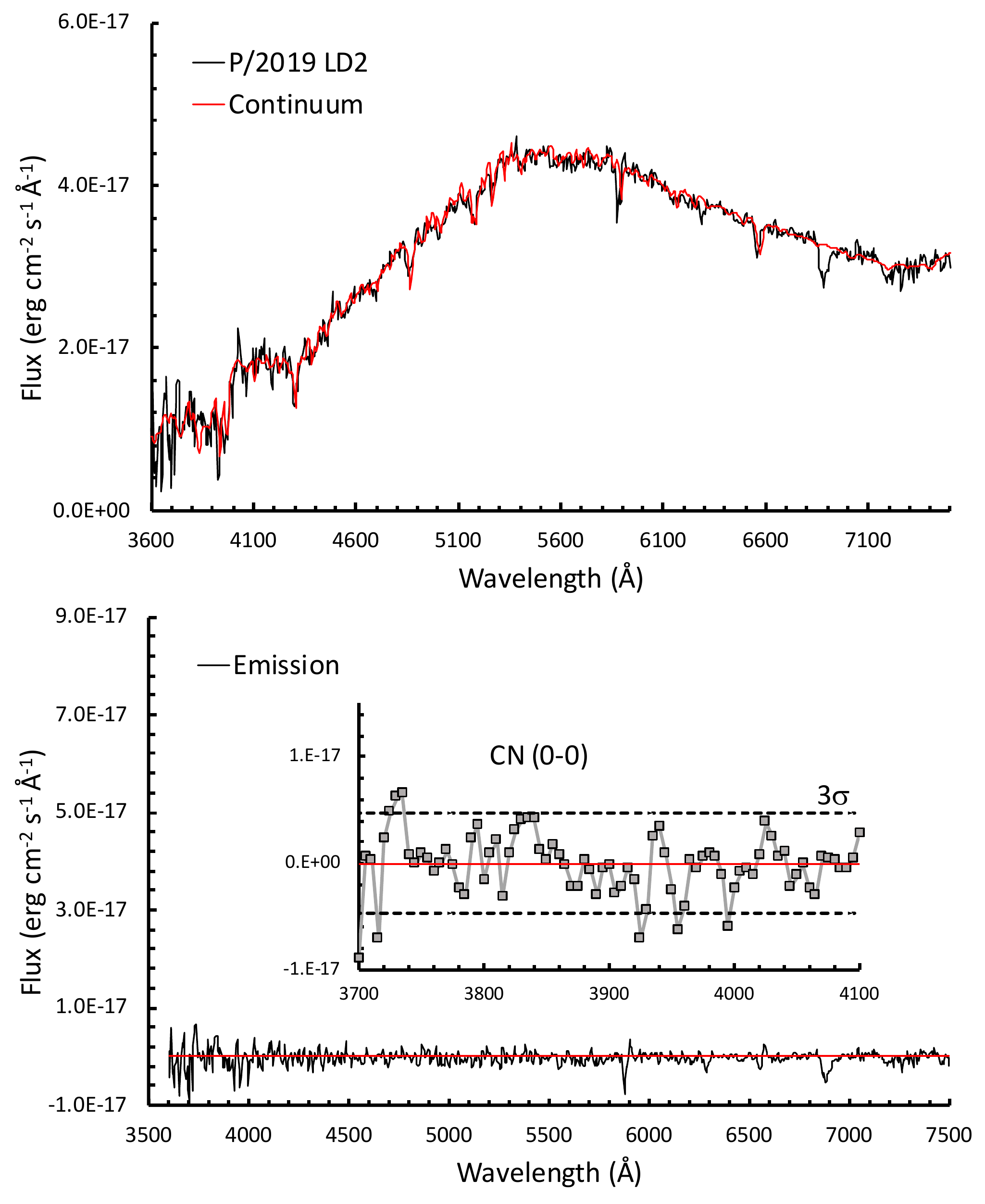}
     \caption{Average of the two individual and flux calibrated spectra of comet P/2019~LD$_{2}$ (ATLAS) obtained with the 10.4~m GTC (upper panel). The spectrum of the Sun was used to obtain the continuum shown in red that was then used to remove  the dust reflected spectrum and obtain the emission spectrum (bottom panel). No CN emission band was detected at the 3$\sigma$ level.}
     \label{VisSpecGTC}
\end{figure}

Spectral images were bias and flat-field corrected, using lamp flats. The 2D spectra were background subtracted, wavelength calibrated (using Xe+Ne+HgAr lamps), and flux calibrated using the spectrophotometric standard star Ross~640. The spectra were then extracted and collapsed to one dimension, using an aperture of $\pm$6 pixels centered at the maximum of the intensity profile of the comet. The value for the extraction aperture corresponds to the distance from the center to where the intensity decreases to 10\% of the maximum. Finally, both spectra were averaged to obtain the final spectrum (see Fig.~\ref{VisSpecGTC}).

%
%
%
%

The final 2D flux-calibrated spectrum is used to analyze the gas emission of comet  P/2019~LD$_{2}$. In order to visualize any emission band associated with cometary species, we used a visible spectrum of the Sun from the CALSPEC compilation \citep{2014PASP..126..711B} to remove the solar continuum.  We scaled the Sun's spectrum  to account for the redness of the spectrum of the target, and then we subtracted it from the comet spectrum (see Fig. ~\ref{VisSpecGTC}). We did not detect any evidence of CN,  C$_2$  or C$_3$ emission within the 3$\sigma$ level. In particular, there were no signs of the CN (0-0) emission at 3880~{\AA}, which  is usually the strongest emission observed in comets. Nevertheless, we were able to  provide an upper limit to the gas production rate of CN using the same procedure as described in \cite{2020MNRAS.495.2053D}, obtaining $Q$(CN) < (1.4 $\pm$ 0.7) $\times$ 10$^{24}$ mol s$^{-1}$.  We did it by using the two regions that border the CN emission band and fitting a linear continuum that was then subtracted from the spectrum. The band flux was measured and converted into column density using the {\em g}-factor in \cite{2010AJ....140..973S} scaled to both the heliocentric distance and velocity of the comet. Then we computed the gas production rate assuming the Haser modeling with the outflow velocity $v_p$ scaled with $r_h$ ($v_p = 0.86 {r_h}^4$ km s$^{-1}$), customary values for the daughter velocity $v_d = 1$ km s$^{-1}$, and scale lengths given in \cite{1995Icar..118..223A}. Theoretical column density profiles for CN were produced for the corresponding set of parameters in the Haser modeling, varying the production rate until the best match between observations and theoretical predictions was achieved.
The non-detection of gas emission in the visible spectrum of LD$_{2}$ is not that surprising as it is very hard to detect them at large heliocentric distances, even for rather large active comets. As an example, in their large compilation of cometary gas production rates, \citet{1995Icar..118..223A} reported a $Q$(CN) = 1.7 $\times$ 10$^{24}$~mol~s$^{-1}$  for comet 74P/Smirnova-Chernykh at 3.56 au. Therefore, it is expected that at the heliocentric distances of LD$_{2}$, the CN production rate, if any, should be below the value 74P at $r_h=3.56$ au.

    \begin{table}
      \centering
        \caption{Observational circumstances of the data presented in this work, obtained in May 2020. Information includes date, airmass (X), heliocentric ($r_h$) and geocentric ($\Delta$) distances, phase angle ($\alpha$), position angle of the projected anti-solar direction ($\theta_{\sun}$), and the position angle of the projected negative heliocentric velocity vector ($\theta_{-V} $). Orbital values are from the JPL HORIZONS system.}\label{observations}
        \begin{tabular}{lcccccc}
           \hline
            Date & X & $r_h$  & $\Delta$ & $\alpha$  & $\theta_{\sun}$ & $\theta_{-V}$\\
                             &       & (au) & (au)        & ($^{\circ}$) & ($^{\circ}$) & ($^{\circ}$) \\
           \hline
           May 16.19 & 1.49& 4.579 & 4.379 & 12.7 & 251.2 & 261.5\\
           May 17.24 & 1.42 & 4.579 & 4.364 & 12.7 & 251.0 & 261.5\\
           \hline
        \end{tabular}
        \end{table}

     \section{Dust tail modeling\label{model}}

    To gain insight into the dust physical properties of
       this object, we used our Monte Carlo dust tail fitting code. 
       In addition to the GTC image obtained on 2020 May 16, and in
order to place stronger constraints on the dust parameters,
we have included in the analysis some of the photometric observations available, namely
the object magnitudes published in the Minor Planet Center (MPEC
2020-K134). Specifically, we considered the reported ATLAS magnitudes
at the time of discovery, on 2019 June 10.4, and those corresponding
to Pan-STARRS1 precovery observations dated as early as 2018 May 21 and 2018 June 10. The ATLAS magnitudes correspond to the 
 orange filter ({\it o}), while the Pan-STARRS magnitudes
correspond to the wide-band purple filter ({\it w}). The measured magnitudes are given in  Table~\ref{magnitudes}. Our model estimates refer to r-Sloan magnitudes, but taking into account the scatter in the measured values and that both measurements refer to wide red bandpasses, we did not apply any photometric correction.

     \begin{table}
      \centering
      \fontsize{8}{12pt}\selectfont
      \tabcolsep 0.15truecm
      \caption{\label{magnitudes}Reported magnitudes of 2019~LD$_{2}$ and
        the corresponding modeled magnitudes from the best-fit Monte
        Carlo dust
        tail model.
              }
      \begin{tabular}{lccc}
       \hline
       Date (UT) & Pan-STARRS 1 & ATLAS-HKO & Model \\
       \hline
       2018 05 21.4 & 21.5 - 21.7 & ---        & 21.2 \\
       2018 06 10.4 & 21.3 - 21.5 & ---        & 21.3 \\
       2019 06 10.4 & --- & 18.18 - 18.55      & 18.5 \\
       \hline
      \end{tabular}
      \end{table}
              

The dust tail
  fitting code has already been described in various papers
  \citep[see, e.g.,][and references therein]{Moreno2016,Moreno2017},
  and it has  proven to be  useful for the characterization of the
  dust environment of normal comets and main belt comets. Briefly, we
  assume that the ejected particles (assumed spherical) describe a
  trajectory that is function of the $\beta$
  parameter. This parameter is the ratio of the solar pressure to 
  the solar gravitational force, and is given by
  $\beta=\frac{C_{pr}Q_{pr}}{\rho_d d}$ \citep{1968ApJ...154..327F},
  where $C_{pr}=1.19\times 10^{-4}$ g cm$^{-2}$, $\rho_d$ is the
  particle density, $d$ is its diameter, and $Q_{pr}$  is the
  scattering efficiency for radiation pressure, which is $Q_{pr}\sim
  1$ for large absorbing particles
  \citep[e.g.,][]{1979Icar...40....1B}.  The tail brightness is
  calculated as the contribution to the brightness of individual
  particles, whose trajectory is computed from the ejection time until
  the time of the observation. Those trajectories depend on $\beta$
  and the terminal velocities. We assume isotropic ejection from a
  spherical nucleus having a certain radius $R_N$. The particles are
  assumed to be distributed in size following a power-law distribution
  with power index $\kappa$. We assume a broad size distribution with
  limiting radii given by  $r_{min}$=10$^{-4}$ cm and $r_{max}$=1 cm,  and {$\kappa$=--2.9. This power-law exponent is a bit larger
    than the typical time-averaged values found in many comets \citep[--4.1
    to --3.0; see][]{2004come.book..565F}, indicating a higher
    relative abundance of larger particles. This value was found as
    the one that best captures the brightness distribution along the
    tail (see Fig.~\ref{Isoph}, right panel).}  The particle density is set nominally to $\rho_d$=1000 kg m$^{-3}$, although we
have also considered a higher density of $\rho_d$=2500 kg m$^{-3}$ \citep[see][]{2019MNRAS.490.2421P}. The geometric albedo is also uncertain. Most centaurs have albedos in the 0.05 to 0.12 range \citep{2020tnss.book..307P},  we thus set $p_v$=0.07 as a typical value. A linear phase function coefficient of 0.03~mag~deg$^{-1}$ is further assumed. For the nucleus, we also assumed $p_V$=0.07, and the same density ($\rho_N$) as for the
particles (i.e., $\rho_N$=$\rho_d$), but a steeper linear phase coefficient of 0.047~mag~deg$^{-1}$, based on the precise estimates for the nucleus of comet 67P from Rosetta/OSIRIS measurements \citep{2015A&A...583A..30F}.

As in our previous works \citep[see,
      e.g.,][]{2019A&A...624L..14M}, the terminal velocities are
parameterized as $v=v_0 \beta^{\gamma}$ , where $v_0$ is a
time-independent speed and the constant $\gamma$ controls the size
dependence of the speed. For activity driven by ice sublimation,
hydrodynamical models predict $\gamma\sim$0.5. However, in situ
measurements of individual dust particles by Rosetta/OSIRIS and GIADA
in the vicinity of comet 67P by \cite{2015Sci...347a3905R}  show no
dependence of particle speed on size at large heliocentric distances (i.e., $\gamma\sim$0). Given the lack of further information, we left
$\gamma$ and  $v_0$ as free parameters of the model. The
remaining model parameters refer to the dust loss rate
distribution. This function is parameterized by a Gaussian function
with peak dust loss rate $(dM/dt)_0$, full width at half maximum
(FWHM), and time of the peak loss rate relative to the GTC observation $t_0$. In order to perform an adequate comparison with the observed tail, the modeled tails are convolved with a Gaussian function with a full width at half maximum equal to the prevailing seeing conditions during the night ($\sim$1.5 \arcsec).

\begin{figure}
   \centering
   \includegraphics[angle=-0,width=\linewidth]{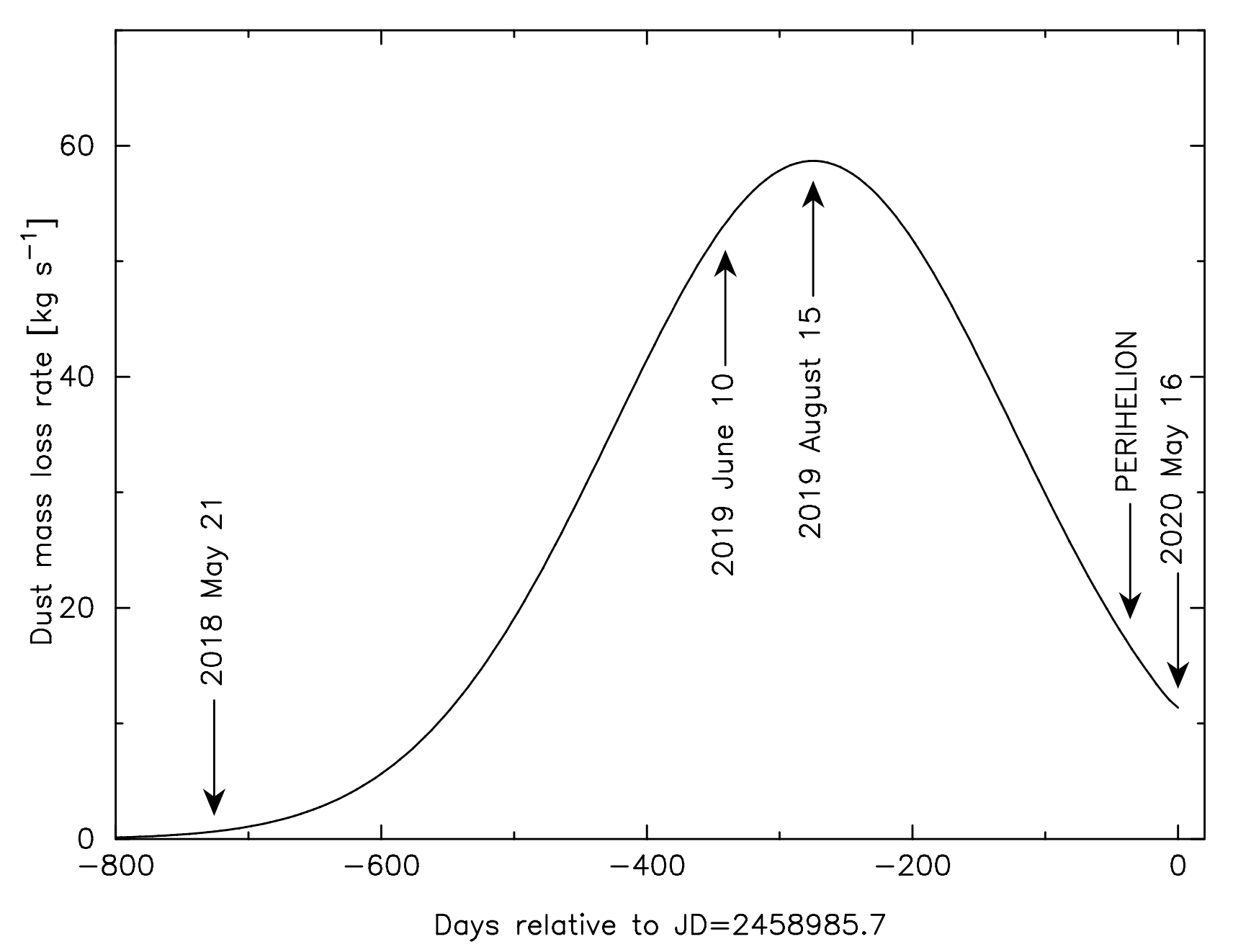}
     \caption{Time evolution of the modeled dust loss rate from 2019~LD$_{2}$ relative to the GTC observation time (2020 May 16.19). The
       perihelion date and the peak emission time are indicated, as
       well as the time of the ATLAS discovery (2019 June 10) and the
       precovery observations from Pan-STARRS 1 (2018 May 21).}
         \label{DustLossRate}
   \end{figure}

\begin{figure*}
   \centering
   \includegraphics[angle=-0,width=\hsize]{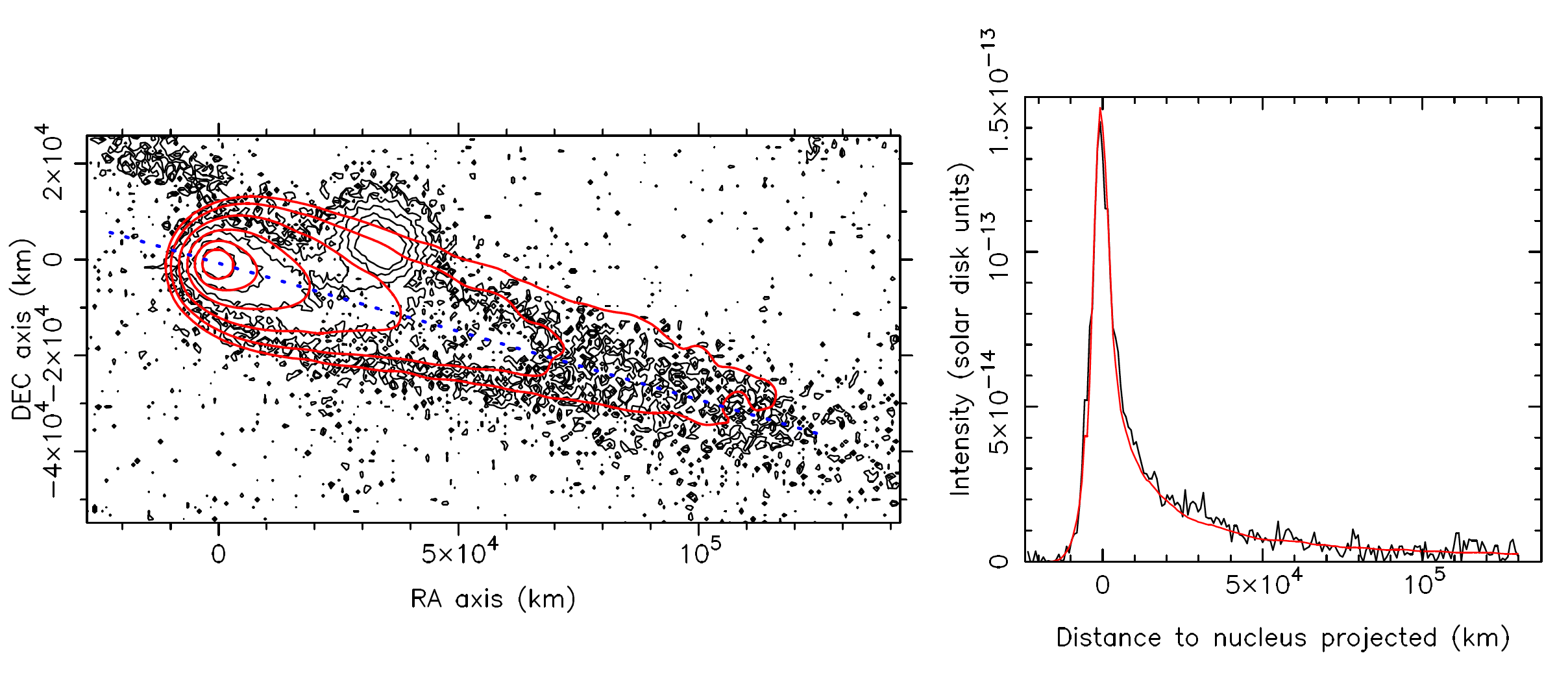}
     \caption{Fitting of the dust model with the observations. Left panel: Comparison of the observed (black
        contours) and modeled (red contours) tail brightness
        isophotes. The innermost contour corresponds to
        8$\times$10$^{-14}$ solar disk units, and the brightness
        decreases in factors of two outward.  One solar
            disk intensity unit corresponds to 2.4$\times$10$^6$
            erg s$^{-1}$ cm$^{-2}$ sr$^{-1}$ \AA$^{-1}$. The images
        are rotated to the conventional orientation (north up, east to the left). The 
        $x$- and $y$-axes are labeled in kilometers projected on the sky at the
        object distance. The blob observed to the right of the comet optocenter is a bright star. Right panel: Comparison of observed (black line)
        and modeled (red line) tail brightness along the direction
        described by the blue dotted line in the left panel.}
         \label{Isoph}
   \end{figure*}

    One of the parameters that influence the innermost
    isophote levels in the computed tails is the assumed nuclear
    radius. This is best guessed from the Pan-STARRS 1 precovery
    observations owing to its faintness, which suggest  
    little or no activity. Assuming complete inactivity, for a nuclear radius of $R_n$=3 km we found $r'$=21.3 for the assumed albedo parameters, which is  in line with the measured magnitudes (see Table~\ref{magnitudes}). The remaining best-fit parameters are found by minimizing the squared sum of the differences between the modeled and measured tail brightness 
 for the GTC image, and the squared sum of the
    differences between synthetic and measured magnitudes at the dates
    given in Table~\ref{magnitudes}. To do 
    this, at each iteration of the model we computed synthetic tail
    images at the corresponding dates, and calculated the r-Sloan synthetic magnitudes from them.  For the GTC image on 2020 May 16, the region of    the CCD affected by the strong 
    brightness by the field star located near RA=35000 km, Dec=5000 km
     (see Figure 5, left panel) is avoided in the fitting
    procedure. The best fit was found by the downhill simplex
method (Nelder and Mead, 1965), resulting in $v_0$=0.8 m s$^{-1}$,
$\gamma$=0.04, $(dM/dt)_0$=60 kg s$^{-1}$,
$t_0$=275 days, and
FWHM=354 days. This implies a total dust mass loss of
1.9$\times$10$^9$ 
kg since the start of the dust emission till the date of
observation. With respect to the GTC
image, a comparison of the isophote fields of the observation
and the model is depicted in Fig.~\ref{Isoph}, left panel. The good
agreement between the brightness along the tail for the observation
and modeled images is shown in the right panel of Fig.~\ref{Isoph}. On
the other hand, the modeled values of the magnitudes agree well with
 the discovery and the precovery observations (see
Table~\ref{magnitudes}). The time evolution of the dust loss rate is
depicted in Fig.~\ref{DustLossRate} where some relevant dates are
indicated. As can be seen, the model predicts that the object was
already active, although with 
a very limited dust production of $\sim$0.6 kg s$^{-1}$, at the time
of the precovery Pan-STARRS 1 observations on 2018 May 21, and
continues to be active as of 
the current epoch (2020 May 16) {(at 11 kg s$^{-1}$)}, spanning
some two years of continuous activity. The maximum level of
    activity of 60 kg s$^{-1}$ is lower than that found for other centaurs at comparable heliocentric distances. Thus, 
    \cite{2006A&A...460..935M} found 100 kg s$^{-1}$ for P/2004 A1 (LONEOS), while 
    \cite{1992Natur.359...42F} and 
    \cite{2009ApJS..183...33M} found 300-900 kg s$^{-1}$ for
    29P/Schwassmann-Wachmann. For 174P/Echeclus, values of 20-40 kg
    s$^{-1}$ have been reported \citep{2008PASP..120..393B}, but at
    much larger heliocentric distance (13 au). Our reported maximum
    production rate is, however, much higher than typical JFCs at the
    same heliocentric distance. For instance, for
    67P/Churyumov-Gerasimenko at 4.5 au, the dust production rate is
    estimated at less than 1 kg s$^{-1}$ 
    \citep{2017MNRAS.469S.186M}. Regarding particle sizes, and as
    described above, we found a
    power-law exponent of $\kappa$=--2.9, which implies the presence of a
     higher amount of large particles compared with those
    usually found in 
    most comets. This is in line with the large particle sizes
    estimated for centaurs P/2004 A1 (LONEOS)
    \citep{2006A&A...460..935M} and 174P/Echeclus
    \citep{2008PASP..120..393B}.

The derived particle speeds are only weakly dependent on size
($\gamma$=0.04), as  was found in situ for comet 67P at 3.7~au \citep{2015Sci...347a3905R}. The speed ($\sim$0.8 m s$^{-1}$)   corresponds to the escape velocity of a $R_N$=1.1 km object with the assumed nominal density of $\rho_N$=1000 kg m$^{-3}$. This is smaller than the radius estimated above, ($R_N$=3 km), but it is in line with it when taking into account the uncertainties on the model parameters, particularly  density and albedo. 

As stated above, we also ran the model considering a higher density for both the nucleus and the particles of 2500 kg m$^{-3}$. In that case, a fit of
similar quality to that depicted in Fig.~\ref{Isoph} was found, but with a steeper size distribution function ($\kappa$=-3.2) and  slightly smaller terminal velocities ($v_0$=0.75 m~s$^{-1}$), the remaining model parameters being the same in both cases.




\section{Implications of the observed activity\label{discusion}}

At the time of the discovery of LD$_{2}$ 
on June 10.4, 2019, the model in Section~\ref{model} shows a dust emission of 
$\sim$50 kg s$^{-1}$, which increases to its  maximum value ($dM/dt$=60 kg
s$^{-1}$) on 2019 August 15 (i.e., roughly eight months before
perihelion), and then decreases again to a  
dust loss rate of 11 kg s$^{-1}$ on 2020 May 16. The fact that this object   became
active when approaching perihelion in the present orbit after the
latest Jupiter encounter, as well as the long-lasting character of the
emission pattern, strongly suggest a thermally driven process as the mechanism responsible for the activity, but the specific mechanism at play in this case is unknown.  

During the activity period the  r$_h$ ranging from 4.5 to 5~au, and CO and other volatiles might be playing a role 
\citep[see][for a review on the activity of distant
  objects]{2017PASP..129c1001W}. However, if CO 
ice sublimation were the dominant driver, and owing to the high
volatility of CO ice, we should have observed a more prominent
coma around the object at the time of Pan-STARRS precovery images on
2018 May 21 (at $r_h\sim$5 au),  which is not the case. This is in
line with the fact that most centaurs do not show activity at large
heliocentric distances, so   their activity cannot be associated
solely with the strong volatility of CO ice \citep{2009AJ....137.4296J}. On the other hand, crystallization of amorphous water ice has been proposed as a plausible mechanism to trigger the outburst activity in centaurs \citep{2009AJ....137.4296J} and distant comets \citep{1992A&A...258L...9P, 2002EM&P...90..217C}, the most paradigmatic case being comet 29P/Schwassmann-Wachmann \citep[see][for a discussion]{2014AN....335..124G}. However, this mechanism is obviously linked to the presence of amorphous ice in the nuclei of such objects. Observations of comet 67P with the Rosetta/ROSINA instrument show that the outgassing pattern is not consistent with the presence of amorphous ice \citep{2016SciA....2E1781L}. Instead, these observations show that the nucleus contains crystalline water ice and clathrates. The activity observed in LD$_{2}$, with a peak well before perihelion, could then be associated with the sublimation of such ices, being exposed at a certain portion of the orbit according to the object's seasons. Interestingly, as in the case of many other centaurs, the activity occurs several months before perihelion \citep{2009AJ....137.4296J}. However, we should also consider that the current level of activity may have been triggered by a sub-catastrophic collision with a smaller body that exposed fresh volatiles from layers below the mantle. Objects moving along co-orbital or nearly co-orbital paths face an increased risk of collisions (see, e.g., \citealt{2018MNRAS.473.3434D}). During the last 3000~yr the only other close encounters with Jupiter inside the Hill radius might have taken place nearly 170~yr ago (at perhaps $\sim$0.06~au), 370~yr ago (at $\sim$0.17~au), and 2642~yr ago (at $\sim$0.27~au), for the nominal orbital evolution in Table~\ref{elements} (see also Fig.~\ref{STevolution} and the discussion in the next section regarding the predictability horizon of these calculations).

\section{Past, present, and future dynamical evolution\label{evolution}}
   The assessment of the dynamical evolution of LD$_{2}$ requires the analysis of an extensive sample of $N$-body 
   simulations. In this work we   use the approach discussed in \citet{2019MNRAS.489..951D} and \citet{2019A&A...625A.133L}. 
   The calculations were performed using the Hermite integration scheme described by \citet{1991ApJ...369..200M} and 
   implemented by \citet{2003gnbs.book.....A}. The standard version of this direct $N$-body code is publicly available from the 
   web site of the Institute of Astronomy of the University of Cambridge.\footnote{http://www.ast.cam.ac.uk/$\sim$sverre/web/pages/nbody.htm} 
   Relative errors in the total energy for the longest integrations presented here are as low as $2\times10^{-12}$ or lower; for 
   the shorter integrations in Fig.~\ref{STevolution} the relative errors in the total energy are always below $5\times10^{-17}$. 
   These values are as good as those in Fig.~4 of \citet{2015MNRAS.452..376R} or better. The relative error in the total angular 
   momentum is several orders of magnitude smaller. As pointed out by \citet{2012MNRAS.427..728D}, the results from this code 
   compare well with those from \citet{2011A&A...532A..89L} among others. In order to generate the initial conditions (control 
   orbits or clones) used in our calculations, we used the orbit determination in Table~\ref{elements}, which is the most 
   recent one (as of  January 27, 2021) and was   released by the Jet Propulsion Laboratory's Solar System Dynamics Group 
   Small-Body Database (JPL's SSDG SBDB, \citealt{2015IAUGA..2256293G})\footnote{\url{https://ssd.jpl.nasa.gov/sbdb.cgi}}.  
%
%
     \begin{table}
      \centering
      \fontsize{8}{12pt}\selectfont
      \tabcolsep 0.15truecm
      \caption{\label{elements}Values of the heliocentric Keplerian orbital elements of P/2019~LD$_{2}$ (ATLAS) and their
               respective 1$\sigma$ uncertainties.
              }
      \begin{tabular}{lcc}
       \hline
        Orbital parameter                                 &   & value$\pm$1$\sigma$ uncertainty \\
       \hline
        Semi-major axis, $a$ (au)                   & = &    5.29537$\pm$0.00004           \\
        Eccentricity, $e$                                 & = &    0.135461$\pm$0.000005         \\
        Inclination, $i$ (\degr)                          & = &   11.55025$\pm$0.00002           \\
        Longitude of the ascending node, $\Omega$ (\degr) & = &  179.75624$\pm$0.00012           \\
        Argument of perihelion, $\omega$ (\degr)          & = &  123.4395$\pm$0.0013             \\
        Mean anomaly, $M$ (\degr)                         & = &    3.1149$\pm$0.0010             \\
        Perihelion distance, $q$ (au)               & = &    4.578053$\pm$0.000008        \\
        Aphelion distance, $Q$ (au)                 & = &    6.01269$\pm$0.00004           \\
        Absolute magnitude, $H$ (mag)                     & = &  12.2$\pm$0.8                         \\
       \hline
      \end{tabular}
      \tablefoot{The orbit determination   refers to epoch JD  2458988.5 (2020 May 19.0) TDB (Barycentric Dynamical Time,
                 J2000.0 ecliptic and equinox). Source: JPL SBDB (solution date,  2021 Jan 14 17:13:11 PST).
                }
     \end{table}
%
%
%
%
     \begin{table}
      \centering
      \fontsize{8}{12pt}\selectfont
      \tabcolsep 0.15truecm
      \caption{\label{vector}Barycentric Cartesian state vector of P/2019~LD$_2$ (ATLAS): components and associated 1$\sigma$ 
               uncertainties.
              }
      \begin{tabular}{ccccc}
       \hline
        Component                         &   &    value$\pm$1$\sigma$ uncertainty                                \\
       \hline
        $X$ (au)                    & = &     2.853532367303273$\times10^{+0}$$\pm$6.29871075$\times10^{-6}$ \\
        $Y$ (au)                    & = &  $-$3.500014237066383$\times10^{+0}$$\pm$6.71754191$\times10^{-6}$ \\
        $Z$ (au)                    & = &     7.142827533391973$\times10^{-1}$$\pm$1.80171333$\times10^{-6}$ \\
        $V_X$ (au/d)                & = &     6.741873338175183$\times10^{-3}$$\pm$2.06450193$\times10^{-8}$ \\
        $V_Y$ (au/d)                & = &     5.156384750948366$\times10^{-3}$$\pm$1.57624641$\times10^{-8}$ \\
        $V_Z$ (au/d)                & = &  $-$1.060557430849704$\times10^{-3}$$\pm$5.38639958$\times10^{-9}$ \\
       \hline
      \end{tabular}
      \tablefoot{ Standard epoch 2459000.5 (2020 May 31.0). Source: JPL's SBDB.
                }
     \end{table}
%
%

   \subsection{Current dynamical status}
      Comet LD$_{2}$ was initially classified as a Jupiter trojan even though the ephemerides showed that it had 
      experienced a close encounter with Jupiter on 2017 February 17, at  0.092~au, well inside the Hill radius of the planet (0.338~au). \citet{2020RNAAS...4...74K}  used an early orbit determination, less precise than the one considered here 
      (123 observations spanning 704~d versus 555 observations spanning 960~d for the orbit determination; see  Table~\ref{elements})
      to conclude that LD$_{2}$ is an active  centaur instead of a Jupiter trojan.  \citet{2021Icar..35414019H} arrived
      at similar conclusions: it  was a centaur prior to July 2018, then a Jovian co-orbital, before returning to centaur after 
      February 2028. They predicted that it will become a JFC after February 2063. The orbit determination used by 
      \citet{2021Icar..35414019H} included 168 observations for a data-arc span of 741~d; therefore, it is also less precise than 
      the orbit investigated here. \citet{2020ApJ...904L..20S} used the same orbit determination considered by \citet{2020RNAAS...4...74K} 
      to conclude that the object will become a member of the JFC dynamical group after 2063.  
      
\begin{figure}
        \centering
        \includegraphics[width=9cm]{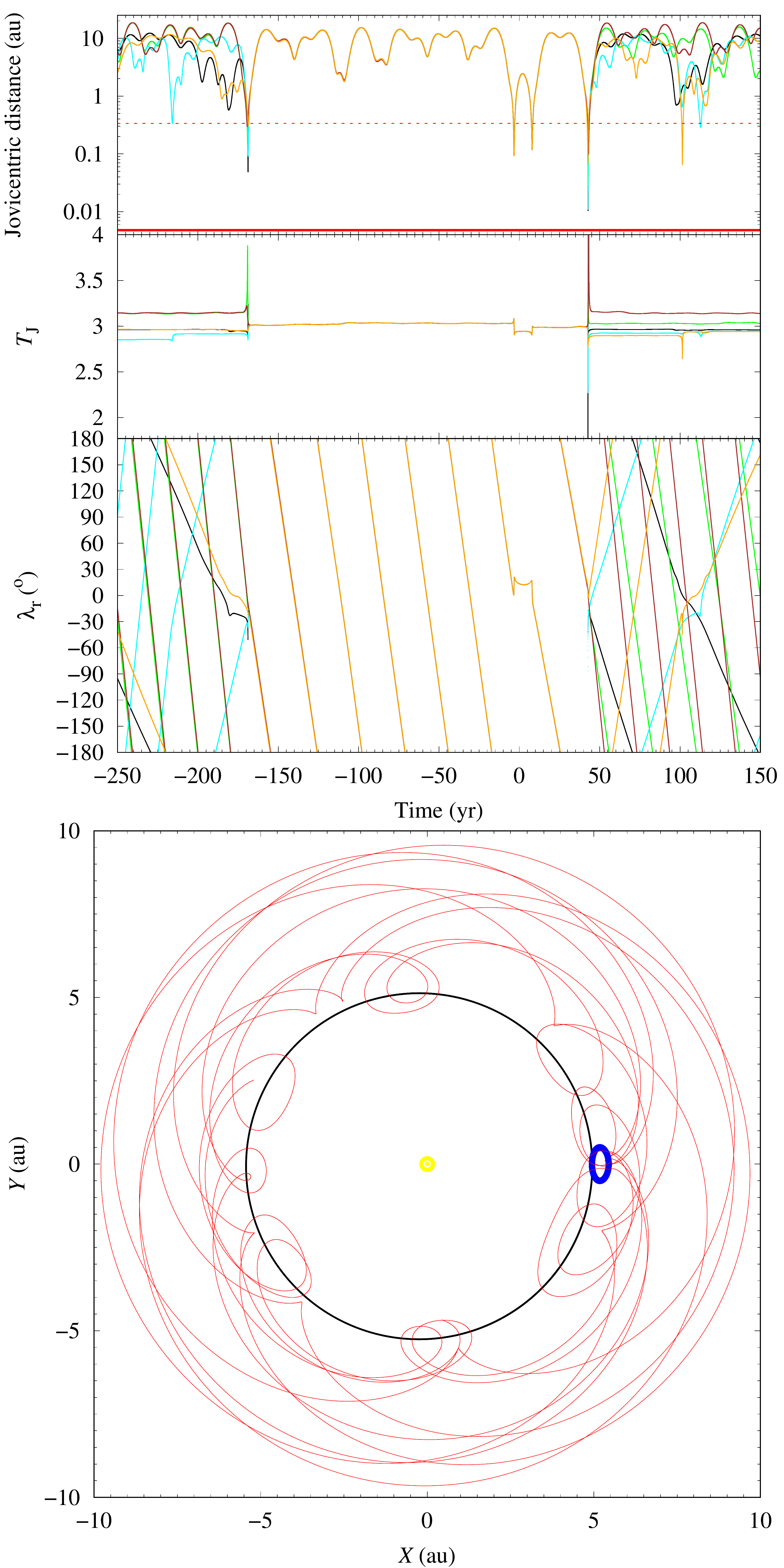}
        \caption{ Short-term evolution of comet P/2019~LD$_{2}$ (ATLAS). The top three panels show the evolution of some 
                 representative parameters   computed from the orbit determination in Table~\ref{elements} (in black) and 
                 relevant control orbits with Cartesian vectors separated $+$3$\sigma$ (in green), $-$3$\sigma$ (in cyan), 
                 $+$9$\sigma$ (in brown), and $-$9$\sigma$ (in orange) from the nominal values in Table~\ref{vector}. The top 
                 panel displays the evolution of the distance from Jupiter to the comet; the red dashed line indicates the Hill 
                 radius of Jupiter at 0.338~au, and the red thick line indicates a distance of 10 Jovian radii. The second panel shows the variation over time of the value of the Tisserand parameter with respect to Jupiter 
                 $T_{\rm J}$ for the same set of reference orbits. The third  panel shows the evolution of the resonant 
                 angle $\lambda_{\rm r}$ for the same sample of control orbits. The bottom panel displays the trajectory followed 
                 by P/2019~LD$_{2}$ in a frame of reference centered at the Sun and rotating with Jupiter, projected on to the 
                 ecliptic plane. The diagram also includes the orbit of Jupiter, its position at (5.2, 0)~au, and the Sun at 
                 (0, 0)~au. This panel shows the evolution of the nominal orbital solution as in Table \ref{elements}. The zero 
                 time instant in the top three panels corresponds to epoch JD 2459000.5 TDB. The blue ellipse in the bottom panel 
                 is described by Jupiter that has a small but non-zero value of the orbital eccentricity. The time span plotted in 
                 the bottom panel is the same as shown in the previous panels.
                }
        \label{STevolution}
     \end{figure}

       Figure~\ref{STevolution} shows the evolution of representative control orbits with Cartesian vectors separated $\pm$3$\sigma$ 
      and $\pm$9$\sigma$ from the nominal values in Table~\ref{vector}.  Figure~\ref{STevolution}, third panel, 
      confirms that LD$_{2}$ is not a Jupiter trojan; the evolution of the resonant angle $\lambda_{\rm r}$ (relative mean 
      longitude with respect to Jupiter) does not exhibit an oscillation about $\pm60\deg$ that is the condition to pursue a 
      tadpole orbit in a frame of reference rotating with Jupiter.  Figure~\ref{STevolution},  third and fourth 
      panels, show that its current dynamical status is similar to that of a quasi-satellite (for about 11~yr) in which the minor 
      body seems to orbit around the planet, although it is not gravitationally bound to it.  Figure~\ref{STevolution}, first, second and 
      third panels, show that the short-term orbital evolution into the past of LD$_{2}$ is robust as the 
      object remains within a safe distance from Jupiter (during the last  $\sim$170~yr), well beyond the Hill radius of the 
      planet. In sharp contrast, the future orbital evolution beyond 2063 is highly uncertain due to a very close encounter with 
      Jupiter (see below). In summary, we  conclude that LD$_{2}$ is now an ephemeral co-orbital of Jupiter, following 
      what looks like a short arc of a quasi-satellite cycle that started in 2017 and will end in 2028. A number of less unstable 
      Jovian quasi-satellites have already been documented (see, e.g., \citealt{2012AcA....62..113W,2016MNRAS.462.3344D}); 
      therefore, the current dynamical status of LD$_{2}$ is not at all surprising. 
%
%
%
%

      Figure~\ref{STevolution} indicates that the predictability horizon as defined originally by \citet{1986RSPSA.407...35L} 
      for P/2019~LD$_{2}$ spans about 43~yr into the future and about 170~yr into the past. Due to its close encounters with 
      Jupiter, it is not possible to predict the evolution of this object for an arbitrarily long amount of time, its trajectory 
      is essentially unstable. This   also affects the value of the Tisserand parameter ($T_{\rm J}$; see Fig.~\ref{STevolution}, 
      second  panel). Although the present-day value of $T_{\rm J}$ is $<3$ and consistent with JFC membership according to 
      \citet{1997Icar..127...13L}, after 2063, the probability of becoming a centaur is slightly higher than that of remaining as 
      a JFC, but the nominal evolution (in black) is consistent with JFC status.

   \subsection{Future orbital evolution}
      As pointed out above, the orbital evolution of P/2019~LD$_{2}$ a few hundred years into the past and up to a few decades 
      into the future can be precisely predicted. However (as shown in  Fig.~\ref{STevolution}), on  January 23, 2063, this object 
      will experience a very close encounter with Jupiter at about  0.016~au. Considering the current uncertainty of its orbit 
      determination, the  computed range of minimum approach distance excludes that the comet may get as close to Jupiter as ten Jovian 
      radii. This very close and slow flyby strongly affects our ability to make reliable dynamical predictions beyond early 2063. 
      In fact, P/2019~LD$_{2}$ may not survive the close approach in one piece due to the strong tidal forces that will presumably 
      occur during the event (tidal breakup may require an approach to 0.001~au in the case of Jupiter); on the other hand, 
      such a close flyby may even lead to a collision with one of the Jovian moons (see, e.g., \citealt{1993Natur.365..731M}). 
      Assuming that P/2019~LD$_{2}$ survives its flyby in 2063, integrations indicate that its probability of escaping the  
      solar system during the next 0.5~Myr is 0.53$\pm$0.03. In general, comets following trajectories similar to that of 
      P/2019~LD$_{2}$ are expected to collide with either the Sun or one of the planets, or to abandon the solar system,  either 
      reaching the Oort Cloud or venturing into interstellar space within a timescale of a few million years (see, e.g., 
      \citealt{1997Icar..127...13L,2020CeMDA.132...36D}). Our calculations appear to confirm a similar outcome for P/2019~LD$_{2}$.

       For these longer calculations into the future and those into the past discussed in the next section, we  used
      the Monte Carlo using the Covariance Matrix (MCCM) methodology described by \citet{2015MNRAS.453.1288D} in which a Monte 
      Carlo process generates control or clone orbits (500) based on the nominal orbit, but adds random noise on each orbital 
      element by making use of the covariance matrix that was retrieved from JPL's SSDG, Horizons On-Line Ephemeris System. 

   \subsection{Past orbital evolution: Possible origin}
      Comet P/2019~LD$_{2}$ has a current value of the Tisserand parameter of 2.94; therefore, and following 
      \citet{1997Icar..127...13L}, it is a JFC. These comets move in very unstable orbits as they experience slow close encounters 
      with Jupiter;  although originally   thought to have an origin in the trans-Neptunian belt (see, e.g., 
      \citealt{1980MNRAS.192..481F,1997Icar..127...13L}), it is now widely assumed that this population has its source in the 
      scattered belt (see, e.g., \citealt{2009Icar..203..140D,2015A&A...573A.102B}). We  performed integrations backward in 
      time  using MCCM to generate control orbits to find that the probability of this comet having been captured from 
      interstellar space during the last 0.5~Myr is  0.49$\pm$0.02 (average and standard deviation).  This probability 
      increased to 0.67$\pm$0.06 for integrations backward in time for 1~Myr, 0.83$\pm$0.06 for 3~Myr, and to 0.91$\pm$0.09 for 
      5~Myr integrations.  The most simple interpretation of these results is that P/2019~LD$_{2}$ almost certainly arrived 
      from interstellar space during the last few million years. It is therefore a dynamically young object, not an object that 
      has remained in its present trajectory since the formation of the solar system. For this reason, the situation described 
      here is very different from that discussed by \citet{2020MNRAS.497L..46M}, who showed that any interstellar planetesimals 
      trapped during the formation of the solar system are highly unlikely to remain with us.

\begin{figure*}
        \centering
         \includegraphics[width=0.49\linewidth]{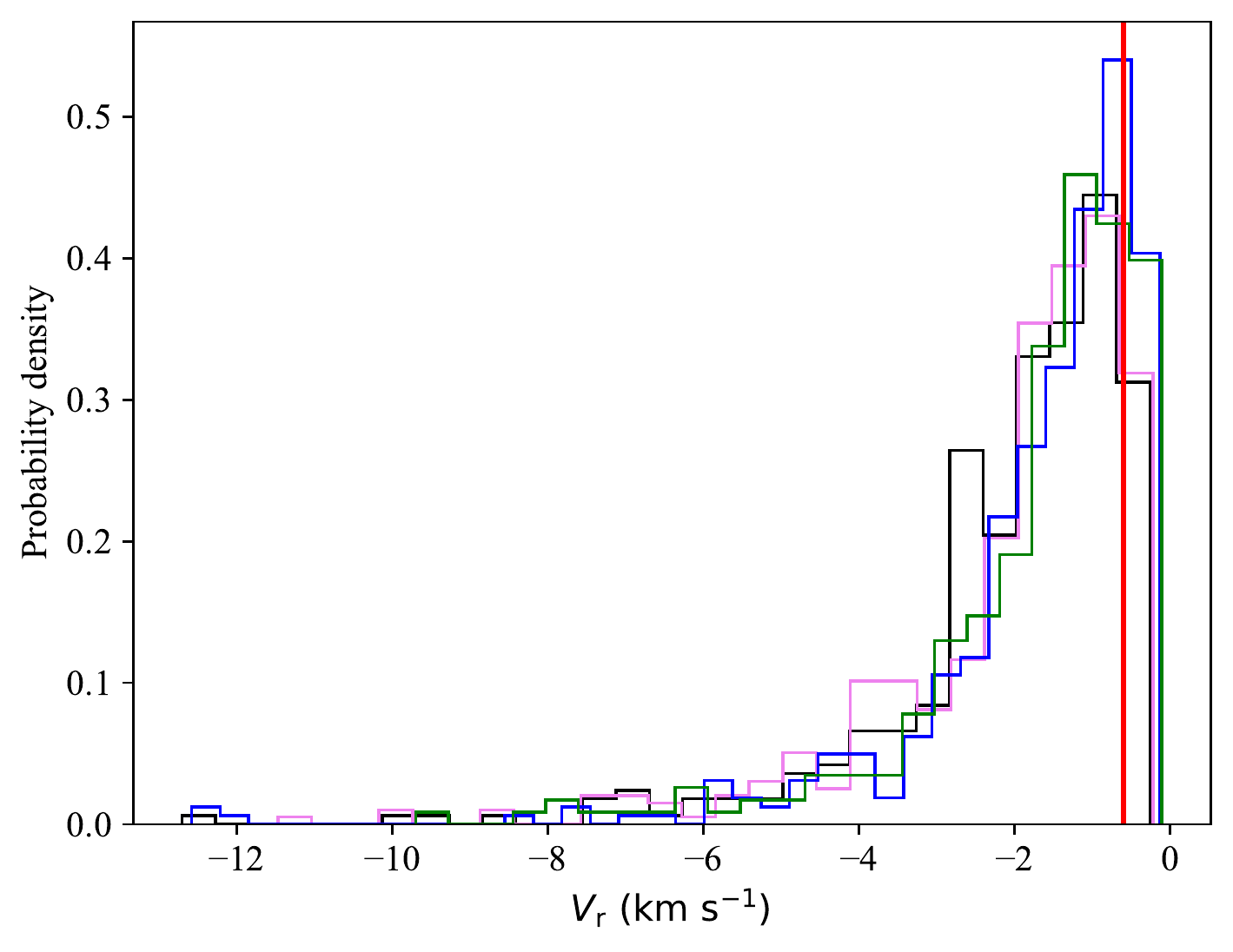}
         \includegraphics[width=0.49\linewidth]{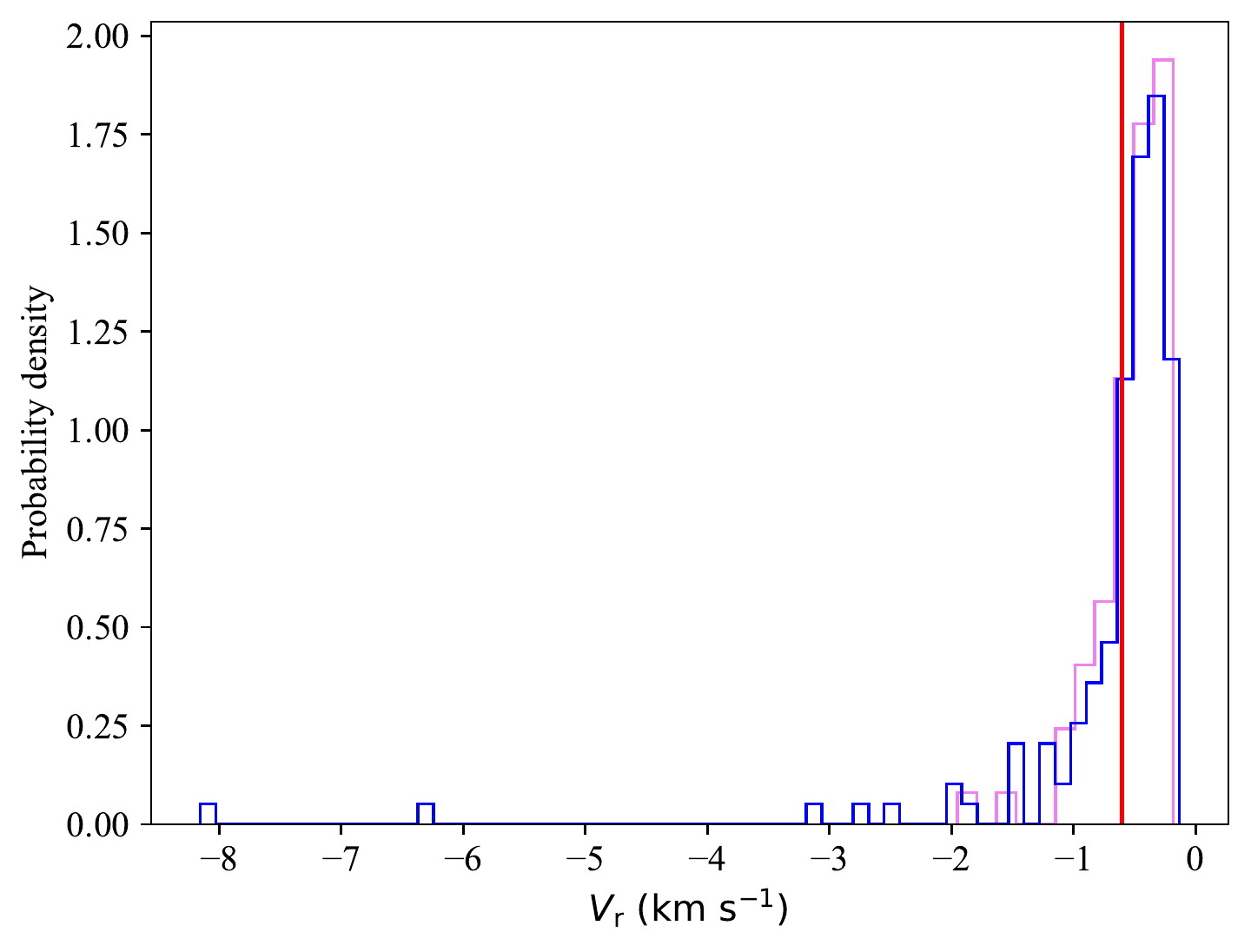}
         \caption{Distributions of inbound velocities for virtual objects  associated with comet 
                  P/2019~LD$_{2}$ (ATLAS) (left) and with comet C/2018~F4 (PANSTARRS) (right). For P/2019~LD$_{2}$ the results 
                  of integrations 0.5~Myr (black), 1~Myr (violet), 3~Myr (blue), and 5~Myr (green) into the past are shown; 
                  for C/2018~F4, the results of the 1~Myr (violet) and 3~Myr (blue) integrations discussed in 
                  \citet{2019RNAAS...3..143D} are displayed. The bins were computed using the Freedman and Diaconis rule 
                  implemented in NumPy \citep{2011CSE....13b..22V}. Counts were used to form a probability 
                  density such that the area under the histogram   sums to one. The vertical red line indicates the value 
                  $-0.6$~km~s$^{-1}$ discussed by \citet{2020MNRAS.493L..59H}.
                 }
         \label{capture}
      \end{figure*}

\begin{figure}
        \centering
         \includegraphics[width=\linewidth]{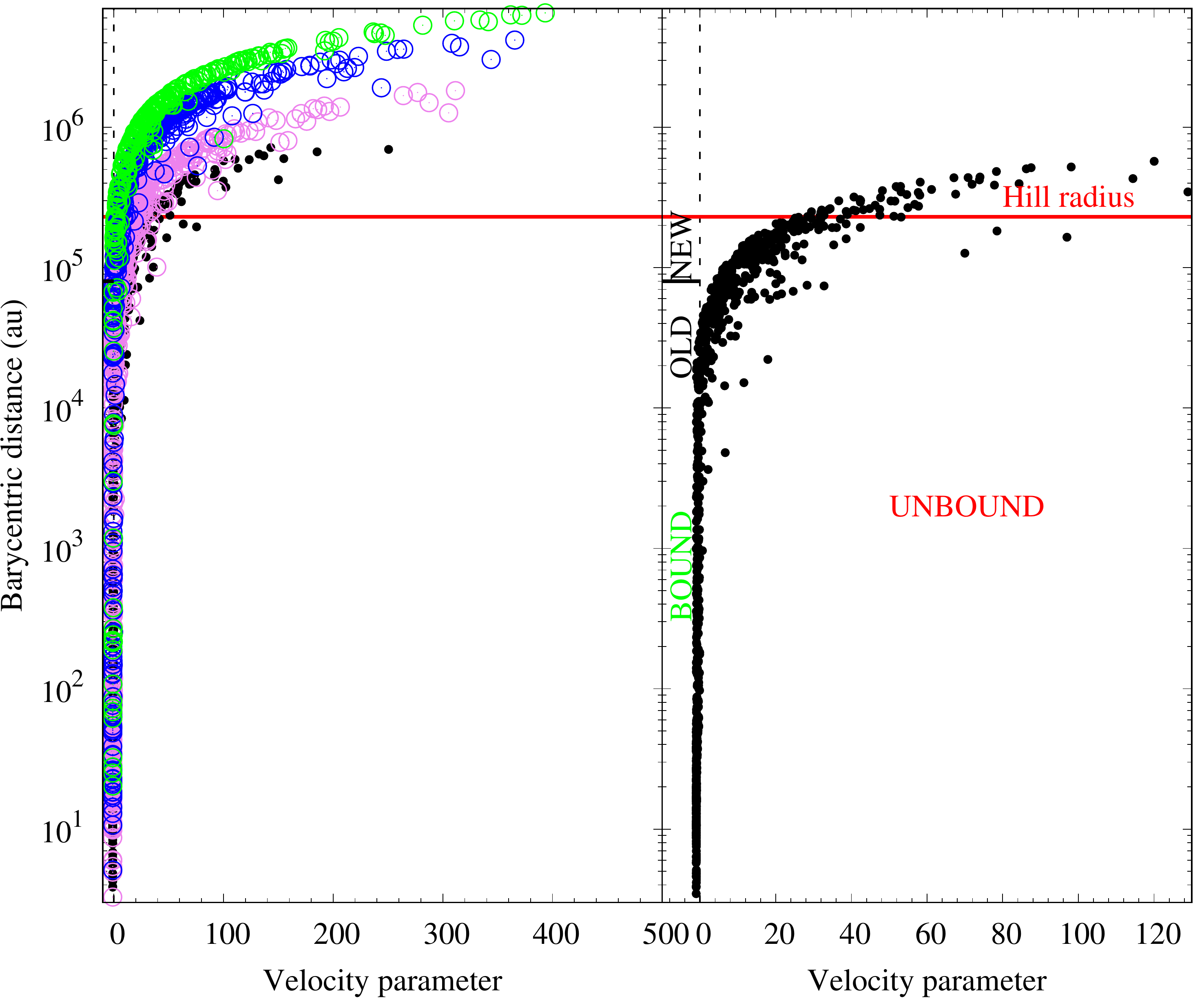}
         \caption{Values of the barycentric distance as a function of the velocity parameter 0.5~Myr (black), 1~Myr (violet), 
                  3~Myr (blue), and 5~Myr (green) into the past (left panel) and future (right panel) for 500 control orbits of P/2019~LD$_{2}$ 
                  (ATLAS). The velocity parameter is the difference between the barycentric and escape velocities at the computed 
                  barycentric distance in units of the escape velocity. Positive values of the velocity parameter identify control 
                  orbits that could be the result of capture (left panel) or lead to escape (right panel). The thick black line 
                  corresponds to the aphelion distance, $a \ (1 + e)$, limiting case $e=1$,  that defines the domain of 
                  dynamically old comets with $a^{-1}>2.5\times10^{-5}$~au$^{-1}$ (see \citealt{2017MNRAS.472.4634K}); the thick 
                  red line indicates  the radius of the Hill sphere of the solar system (e.g., \citealt{1965SvA.....8..787C}).
                 }
         \label{origin}
      \end{figure}

      Figure~\ref{capture}, left panel, shows the distribution of inbound velocities for virtual objects (control orbits, see 
      above) in hyperbolic paths with respect to the barycenter of the solar system 0.5~Myr (black),  1~Myr (violet), 3~Myr 
      (blue), and 5~Myr (green) into the past. The median, and the 16th and 84th percentiles of the velocity distributions are 
       $-1.6_{-1.4}^{+0.9}$~km~s$^{-1}$ (0.5~Myr), $-1.5_{-1.9}^{+0.8}$~km~s$^{-1}$ (1~Myr), $-1.2_{-1.5}^{+0.7}$~km~s$^{-1}$ (3~Myr), 
      and $-1.3_{-1.5}^{+0.8}$~km~s$^{-1}$ (5~Myr). In figure~2 of \citet{2020MNRAS.493L..59H} the distribution of hyperbolic excess velocities for captured 
      interstellar objects from simulations is shown; it exhibits a maximum at about 0.6~km~s$^{-1}$, which  corresponds to an 
      inbound velocity at large distance from the solar system of $\sim$$-0.6$~km~s$^{-1}$. Most virtual interstellar objects 
      associated with P/2019~LD$_{2}$ have inbound velocities at over 1~pc from the Sun close to the most probable value 
      in \citet{2020MNRAS.493L..59H} (see Fig.~\ref{capture}, left panel, green histogram). On the other hand, Fig.~\ref{capture}, 
      right panel, shows the distribution of inbound velocities for the case of hyperbolic comet C/2018~F4 (PANSTARRS) as 
      discussed in \citet{2019RNAAS...3..143D}; for this object  the dynamics and the observational data both strongly suggest 
      that it is a former member of the Oort Cloud \citep{2019A&A...625A.133L}. Figure~\ref{origin} uses data from the same 
      simulations plotted in Fig.~\ref{capture} and further hints at an extrasolar origin for P/2019~LD$_{2}$; most clones are
      found well beyond the radius of the Hill sphere of the solar system 5~Myr ago.  Although an origin outside the 
      solar system for P/2019~LD$_{2}$ seems plausible and reasonably well supported by the available evidence, we also have to 
      admit that the actual sequence of events that led to what is observed today could have been more complex. A former member of the 
      scattered disk may have experienced a very close encounter with one of the giant planets after becoming part of the centaur 
      dynamical class in the relatively recent past, less than 0.5 Myr ago; such an encounter may have produced a fragmentation 
      event induced by the planetary tidal force that was eventually able to form the observed present-day P/2019~LD$_{2}$. The
      feasibility of such events was dramatically confirmed by  comet Shoemaker-Levy~9 in 1992 (see, e.g.,
      \citealt{1993IAUC.5800....1N}). However, the available orbit determination of P/2019~LD$_{2}$ is not robust enough to either 
      confirm or reject this more complex scenario  that seems a priori more likely, taking into account its low orbital 
      inclination.
%
%
%
%
%
%
%
%

   \section{Conclusions\label{conclusions}}

      In this paper we  presented observations of Jupiter's transient co-orbital comet P/2019~LD$_{2}$ (ATLAS) obtained on 2020 May 16 and May 17, using the OSIRIS camera-spectrograph  at the  10.4~m GTC. We used the obtained images and spectra to characterize its overall level of cometary activity. We also carried out direct $N$-body simulations to investigate its orbital evolution. Our conclusions can be summarized as follows:
      \begin{enumerate}[(i)]
        \item  LD$_{2}$ shows a conspicuous coma and tail with a length of about  1\arcmin. 
        \item There is no evidence of CN, C$_2$, or C$_3$ emission within the 3 $\sigma$ level in the acquired spectrum. In particular, there is no sign of the CN (0-0) emission at 3880~{\AA} that it is usually the strongest emission observed in comets. We obtain an upper limit to the CN gas production rate $Q$(CN)<(1.4 $\pm$ 0.7) $\times$ 10$^{24}$~mol~s$^{-1}$.  The non-detection of CN at 4.5 au for a small comet like P/2019~LD$_{2}$ has allowed us to place a rather conservative upper limit. \citet{1995Icar..118..223A} reported a production rate $Q$(CN) = 1.7 $\times$ 10$^{24}$~mol~s$^{-1}$  for Jupiter family comet 74P/Smirnova-Chernykh at 3.56 au. Therefore, it is expected that at the heliocentric distances of LD$_{2}$, the CN production rate, if any, should be below the value 74P at $r_h=3.56$ au. The acquired data does not allow us for a more stringent upper limit.
        \item The comet brightness in a 2.6\arcsec aperture diameter is  $r' = 19.34\pm0.02$. The coma is redder than the Sun with colors  $(g'-r') = 0.78\pm0.03$, $(r'-i') = 0.31\pm0.03$, and $(i'-z')= 0.26\pm0.03$.  
        \item According to our model, the dust emission of
          LD$_{2}$ can be described by a Gaussian with a
          FWHM=354 days, a maximum $(dM/dt)_0$=60 kg s$^{-1}$ reached          on August 15, 2019 ($t_0$=275 days before the observations), which  then decreases again, with a current (2020 May 16)
          dust loss rate of 11~kg~s$^{-1}$.  This implies a total dust
          mass loss of 1.9$\times$10$^9$ kg since the start of the
          dust emission, and very little activity (0.6 kg s$^{-1}$)
          at the time of precovery Pan-STARRS 1 observations in May 2018.       \item The origin of the observed activity is most likely linked to a thermally driven process, associated with sublimation of crystalline water ice and clathrates, either by a seasonal effect or triggered by a collision with a smaller body.  
        \item From the image photometry, we obtained a lower limit for the absolute magnitude $H_g = 13.10\pm0.03$~mag thus an upper limit for the nucleus radius $R_N$ between 5.0 and 8.0~km. With the Monte Carlo dust tail fitting code and considering the precovery Pan-STARRS 1 magnitude data of the nearly bare nucleus, a nuclear radius of $\sim$3 km is derived, and this value is found to be compatible with the GTC dust tail brightness in the near-nucleus region. The derived particle speed of  $\sim$0.8~m~s$^{-1}$ corresponds to the escape velocity from a $R_N$=1.1~km object with the assumed nominal density of $\rho_N$=1000~kg~m$^{-3}$. All results show that LD$_{2}$ is a kilometer-sized object, in the typical size-range of the JFCs.
         \item  LD$_{2}$ is now an ephemeral co-orbital comet of Jupiter, following what looks like a short arc of a quasi-satellite cycle that started in 2017 and will end in 2028. 
        \item  LD$_{2}$ will experience a very close encounter with Jupiter at perhaps 0.016~au on January 23, 2063. If it survives the close approach, its probability of escaping the solar system during the next 0.5~Myr is 0.53$\pm$0.03.
         \item The origin of LD$_{2}$ is still an open question. The probability of this comet having been captured from interstellar space during the last 0.5~Myr is 0.49$\pm$0.02 (average and standard deviation),  0.67$\pm$0.06 during the last 1~Myr, 0.83$\pm$0.06 for 3~Myr, and of 0.91$\pm$0.09 for 5~Myr, suggesting that LD$_{2}$ may be a temporarily captured interstellar comet. However, it cannot be discarded that a very close encounter with one of the giant planets of a former member of the scattered disk may have triggered a fragmentation event that was eventually able to produce the observed present-day LD$_{2}$.
      \end{enumerate}
       Although the physical characterization of this object based on the data presented here can be regarded as robust, we note that reconstructing its past behavior as well as predicting its future dynamical evolution remains very challenging within the context of its current orbit determination. If observations acquired prior to 2018 are made public (they have already been found, as discussed by \citealt{2020arXiv201109993K}), the uncertainty associated with its past and future will be reduced considerably.
       
       After the acceptance of this paper, an initial characterization of LD$_{2}$ was published by \citet{2021AJ....161..116B}. They report an absolute magnitude $H_V = 15.53 \pm 0.05$, which corresponds to a nucleus radius  $R_N$=2~km assuming a value for the albedo $p_V$=0.07 (the one used in our dust model). Although their value is slightly smaller than our $R_N$=3~km determination, it lends further support to our conclusion that LD$_{2}$ is a kilometer-sized object. The colors they report ($(g'-r') = 0.60 \pm 0.03$, $(r'-i') = 0.18 \pm 0.05$, $(i'-z')= 0.01 \pm 0.07$) are also slightly different. As this is an extended object the aperture used is important for comparison purposes. Using the same aperture (equivalent to 10,000 km at the comet distance) we obtain $(g'-r') = 0.73 \pm 0.03$, $(r'-i') = 0.37 \pm 0.03$, $(i'-z')= 0.21 \pm 0.07$ , still redder than Bolin's values.  Based on tail dimensions, they assume the dust tail as being populated by particles with a mean radius of only 400 micrometers. The inferred velocity of $\sim1$ m/s agrees with our findings. However, the total dust loss rate they infer is smaller than ours by about one order of magnitude, which we attribute to the different dust models used, mainly  the consideration of a size distribution in our more realistic Monte Carlo dust tail model. They also did not detect $C_2$ in the spectrum of LD$_{2}$, providing an upper limit for its production rate that is 5.4 times higher than our computed upper limit for the $CN$ production rate. The  $Q$(C$_2$/CN) production rate is lower than 2  for the large majority of observed comets \citep{1995Icar..118..223A} so our spectrum is likely more sensitive to gas production.

   \begin{acknowledgements}
      JdL acknowledges support from MINECO under the 2015 Severo Ochoa Program SEV-2015-0548. FM and LL acknowledge financial support from the State Agency for Research of the Spanish MCIU through the "Center of Excellence Severo Ochoa" award to the Instituto de Astrof\'\i sica de Andaluc\'\i a (SEV-2017-0709). FM also acknowledges financial support from the Spanish Plan Nacional de Astronom\'\i a y Astrof\'\i sica LEONIDAS project RTI2018-095330-B-100, and project P18-RT-1854 from Junta de Andaluc\'\i a. MDeP acknowledges funding from the Preeminent Postdoctoral Program of the University of Central Florida and from the SRI/FSI project "Finding the recipe to cook a primitive small body in the Solar System''. NP-A acknowledges support from SRI/FSI funds through the project "Diggin-up Ice Rocks in the Solar System". This research was partially supported by MINECO under grant ESP2017-87813-R. RdlFM and CdlFM thank S.~J. Aarseth for providing one of the codes used in this research and for comments on the evolution of hyperbolic comets, and A.~I. G\'omez de Castro for providing access to computing facilities. Part of the calculations and the data analysis were completed on the Brigit HPC server of the `Universidad Complutense de Madrid', and we thank S.~Cano Als\'ua for his help during this stage. In preparation of this paper, we made use of the NASA Astrophysics Data System, the ASTRO-PH e-print server, and the MPC data server.
   \end{acknowledgements}

\bibliographystyle{aa} 
\bibliography{38842.bib}

\end{document}